\documentclass[11pt]{article}
\PassOptionsToPackage{switch}{lineno}
\usepackage{acl}
\usepackage{times}
\usepackage{latexsym}
\usepackage{amsmath}
\usepackage[T1]{fontenc}
\usepackage{multirow}
\usepackage[utf8]{inputenc}
\usepackage{amssymb}
\usepackage{booktabs}
\usepackage{microtype}
\usepackage{inconsolata}
\usepackage{graphicx}
\usepackage{tabularx}
\usepackage{pgfplots}
\usepgfplotslibrary{groupplots}
\pgfplotsset{compat=1.18}
\title{ParsVoice: A Large-Scale Multi-Speaker Persian Speech Corpus for Text-to-Speech Synthesis}
\author{Mohammad Javad Ranjbar Kalahroodi$^1$, Heshaam Faili$^1$, Azadeh Shakery$^{1,2}$ \\
	$^1$School of Electrical and Computer Engineering, University of Tehran, Iran \\
	$^2$Institute for Research in Fundamental Sciences (IPM), Tehran, Iran \\
	\texttt{\{mohammadjranjbar, hfaili, shakery\}@ut.ac.ir}}
\begin{document}
	\maketitle
	
\begin{abstract}
	Persian remains substantially underrepresented in open speech-text resources, limiting progress in multi-speaker text-to-speech (TTS), speech-language modelling, and low-resource speech processing. We introduce \textbf{ParsVoice}, the largest publicly available Persian speech-text corpus tailored for training multi-speaker TTS systems, along with a scalable pipeline to construct high-quality speech-text data from long-form audiobook recordings. The pipeline combines a fine-tuned ParsBERT sentence-completion classifier, ASR-based boundary optimization, punctuation restoration, speaker identification, and a multi-dimensional quality assessment that covers both audio and Persian-specific text properties. The resulting release contains a 2,200-hour TTS-ready subset with 1.36 million aligned segments from 1,815 automatically inferred speaker IDs, making it more than 25 times larger than the largest previously available open Persian TTS dataset. To validate the corpus, we fine-tune XTTSv2, a zero-shot multilingual TTS model that operates directly on raw Persian text without phoneme representations. The resulting model achieves a naturalness MOS of 3.6/5 and a speaker-similarity MOS of 4.0/5. ParsVoice, its metadata, and the corpus-construction pipeline are publicly available on Hugging Face\footnote{\url{https://huggingface.co/datasets/MohammadJRanjbar/ParsVoice}} and GitHub\footnote{\url{https://github.com/MohammadJRanjbar/ParsVoice}}, supporting reproducible research on Persian speech synthesis and low-resource speech-language technologies.
\end{abstract}

	\section{Introduction}
	\label{sec:intro}
	The rapid advancement of transformer architectures \cite{vaswani2017attention} and generative models has increased the demand for large-scale, high-quality speech data. Despite this growing need, resource availability remains highly uneven across languages, and Persian remains significantly underrepresented in speech corpora compared to high-resource languages such as English. 
	
\begin{figure}[t]
	\centering
	\includegraphics[width=\columnwidth]{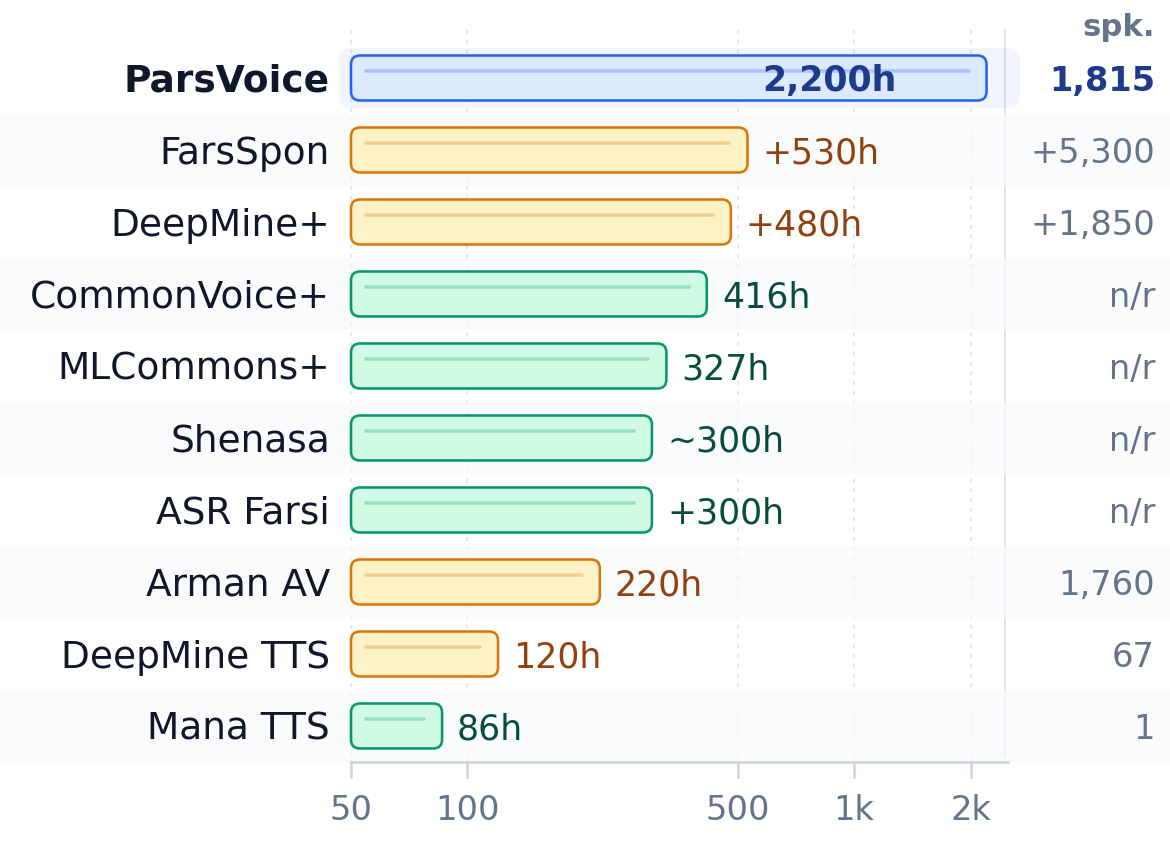}
	\caption{Comparison of Persian speech datasets by duration, in hours of
		audio on a logarithmic scale. Speaker counts are shown at right;
		\textit{n/r} indicates that the speaker count is not reported.}
	\label{fig:dataset_comparison}
\end{figure}
	
	\begin{figure*}[t]
		\centering
		\includegraphics[width=0.95\linewidth]{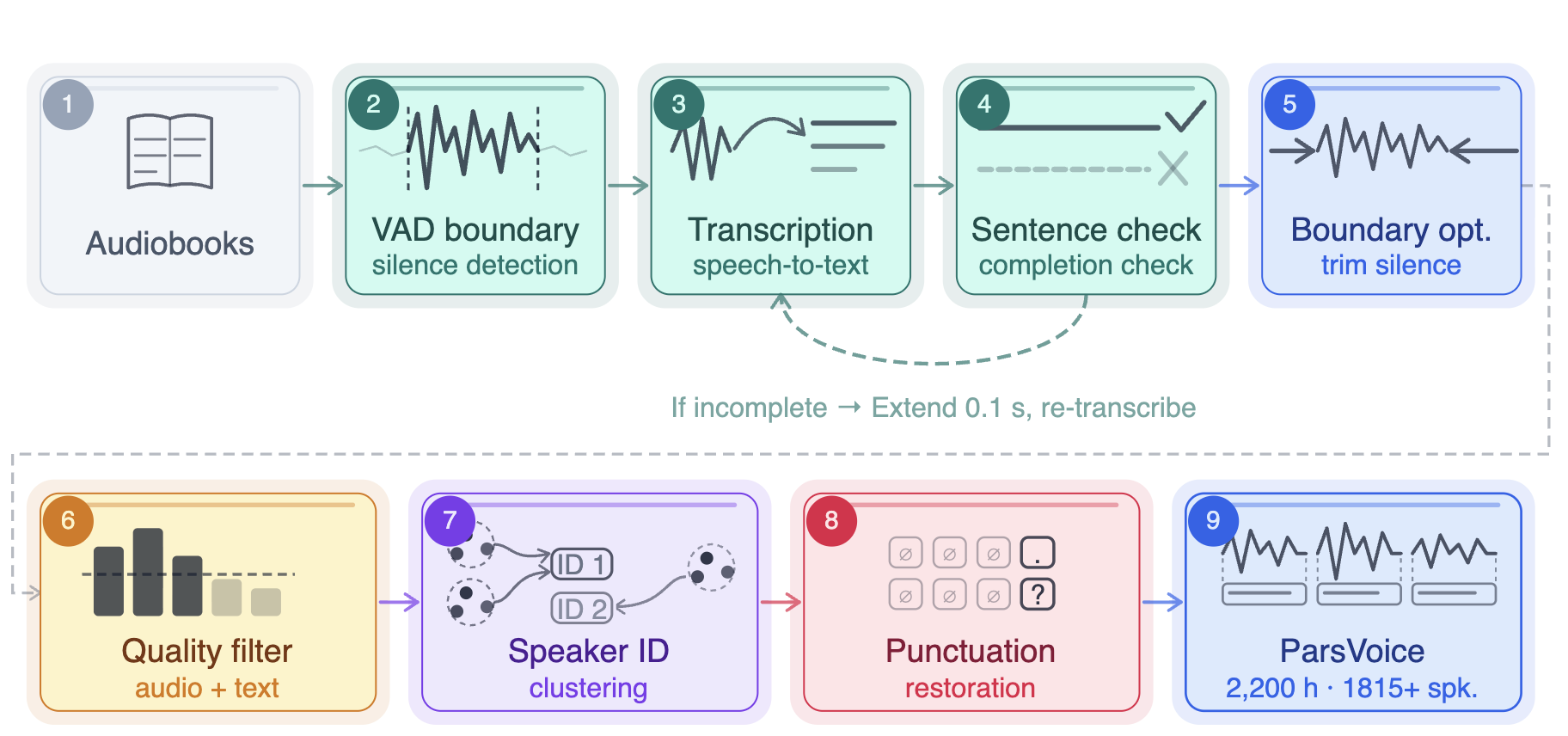}
		\caption{Overview of the ParsVoice corpus construction pipeline. Stages 1--5 handle segmentation and transcription; the dashed arc between stages 3 and 4 shows the iterative extension loop for incomplete sentences. Stages 6--9 handle filtering, speaker identification, punctuation restoration, and final output.}
		\label{fig:pipeline}
	\end{figure*}
	
	This scarcity extends beyond speech: insufficient Persian data affects a broad range of language technologies, including ASR, forced alignment, and optical character recognition (OCR). English benefits from thousands of hours of high-quality labelled speech \cite{panayotov2015librispeech}, mature alignment tools with pretrained acoustic models \cite{mcauliffe17_interspeech}, and well-established OCR benchmarks---resources that are simply unavailable at a comparable scale for Persian. Moreover, Persian speech presents several language-specific challenges: short vowels are systematically omitted in written text, the \textit{Ezafe}---a connective vowel between consecutive words---is never written yet must be pronounced, and a single letter can map to multiple phonemes depending on context \cite{Adibian2025DeepMine}. These properties make raw-text Persian TTS challenging and have motivated many prior systems to rely on explicit phonetic or linguistic preprocessing. This resource gap has slowed development across multiple areas of Persian language processing and widened the digital divide between Persian and better-resourced languages.

	TTS is particularly demanding in its data requirements. Unlike ASR models, which can tolerate noisy, real-world recordings, TTS models require data that satisfies several strict criteria simultaneously: audio must be free of background noise, music, and artifacts surrounding the speech signal, as background noise directly impairs alignment and degrades synthesised speech quality \cite{zen2019libritts}; transcripts must consist of complete, well-punctuated sentences, since punctuation governs prosody modelling and its absence measurably reduces naturalness \cite{zen2019libritts,fetrat2024mana}; speech and text must be precisely aligned at sentence boundaries rather than arbitrary silence points \cite{zen2019libritts}; and the corpus should cover a diverse range of speakers and content domains to enable generalisation across voices and topics \cite{zen2019libritts,nakamura2025ttsops}. Meeting all of these requirements at scale is significantly more difficult and expensive than collecting ASR data, and the challenge is compounded for Persian by the linguistic properties discussed above. Compared with the wide range of English speech resources \cite{panayotov2015librispeech,ito2017lj,veaux2017cstr}, Persian TTS datasets are still limited in scale, speaker diversity, and accessibility. Most are single-speaker and often proprietary, which restricts the development of multi-speaker TTS systems.
	
	To address this gap, we introduce \textbf{ParsVoice}, a large-scale publicly available Persian speech-text corpus designed to provide high-quality training data for Persian TTS and low-resource speech-language research. ParsVoice targets the harder setting of constructing a multi-speaker corpus from long-form audiobook recordings without relying on publicly available reference transcripts. We develop a scalable automated pipeline that combines sentence-aware segmentation, ASR-based transcription, a ParsBERT sentence-completion classifier, ASR-based boundary optimization, speaker identification, punctuation restoration, and Persian-specific audio/text quality assessment. The final TTS-ready release contains 2,200 hours of speech, 1.36 million aligned segments, and 1,815 automatically inferred speaker IDs. In terms of duration, this expands the scale of openly available Persian TTS data by more than a factor of 25 relative to ManaTTS~\cite{fetrat2024mana}.
	
	To validate ParsVoice, we fine-tune XTTSv2~\cite{casanova2024xttsmassivelymultilingualzeroshot}, a zero-shot multilingual TTS model that operates directly on text without explicit phoneme representations. On unseen Persian reference speakers, the adapted model achieves a naturalness MOS of 3.60/5 and a speaker-similarity SMOS of 4.03/5, demonstrating the suitability of ParsVoice for multi-speaker Persian TTS. Figure~\ref{fig:pipeline} provides an overview of the corpus construction pipeline.

	The remainder of the paper is organized as follows. Section~\ref{sec:related} reviews related Persian speech resources and TTS systems, Section~\ref{sec:pipeline} describes the ParsVoice construction pipeline, Section~\ref{sec:corpus} analyzes the resulting corpus, and Section~\ref{sec:tts_eval} evaluates TTS training with ParsVoice. Sections~\ref{sec:conclusion}--\ref{sec:ethics} conclude with the main findings, limitations, and ethical considerations.
		
	\section{Related Work}
	\label{sec:related}
	
	\subsection{Speech Corpora for High-Resource Languages}
	Speech dataset development has been dominated by English resources. LibriSpeech \cite{panayotov2015librispeech} provides 960 hours of read speech from audiobooks across 2,484 speakers, LJSpeech \cite{ito2017lj} offers 24 hours from a single speaker, and VCTK \cite{veaux2017cstr} contains approximately 44 hours across 109 speakers. Multilingual efforts such as Common Voice \cite{ardila2020common}, Multilingual LibriSpeech \cite{pratap2020mls}, and VoxPopuli \cite{wang2021voxpopuli} have expanded coverage to 20+ languages, but remain skewed toward European languages with variable quality across linguistic communities. Many widely spoken non-European languages---including Persian---have received comparatively little attention in these initiatives.
	
	\subsection{Persian Speech Datasets}
	
	Existing Persian speech resources suffer from critical limitations in scale, speaker diversity, and accessibility. Figure~\ref{fig:dataset_comparison} compares ParsVoice with representative Persian speech datasets in terms of duration.
	
	Prior to this work, DeepMine+~\cite{zeinali2019deepmine} was the largest existing Persian speech resource, providing over 480 hours of speech from more than 1,850 speakers; however, it is commercially restricted. Among TTS-specific datasets, DeepMine Multi-TTS~\cite{Adibian2025DeepMine} provides 120 hours across 67 speakers under a restricted license with manually verified transcripts and phoneme-level annotations; ArmanTTS~\cite{shamgholi2023armantts} contains approximately 9 hours from a single speaker; and ManaTTS~\cite{fetrat2024mana} offers 86 hours from a single speaker under an open CC-0 license. Common Voice's Persian portion exists but lacks the audio quality required for TTS training.
	
	\subsection{Persian TTS Systems}
	Existing Persian TTS systems have predominantly relied on explicit phoneme-level intermediate representations, adding pipeline complexity and requiring language-specific lexicons or grapheme-to-phoneme models. ManaTTS~\cite{fetrat2024mana} introduced an end-to-end Persian TTS system built on a Tacotron-style architecture and trained on its accompanying 86-hour single-speaker corpus, with phoneme sequences derived from a rule-based front-end. ArmanTTS~\cite{shamgholi2023armantts} similarly targets single-speaker synthesis from a 9-hour corpus using a standard sequence-to-sequence TTS pipeline with phoneme inputs. The most closely related prior work is DeepMine Multi-TTS~\cite{Adibian2025DeepMine}, which trains Tacotron2 and FastSpeech2 models on their dataset, using phoneme sequences rather than raw Persian text as input.
	For corpus validation, we fine-tune XTTSv2~\cite{casanova2024xttsmassivelymultilingualzeroshot}, which operates \emph{directly} on raw Persian text without any phoneme transcription step. This choice avoids the explicit phonetic front-end used by several previous Persian TTS systems. A quantitative comparison of MOS scores is presented in Section~\ref{sec:tts_eval}.
	
	\section{ParsVoice}
	\label{sec:pipeline}
	ParsVoice is constructed using a scalable pipeline that transforms raw audiobook recordings into a structured, high-quality speech-text corpus. The pipeline addresses three key challenges in Persian speech data creation: preserving sentence integrity, ensuring accurate audio-text alignment, and processing thousands of hours of speech at scale.
	
	\subsection{Data Collection and Source Selection}
	We selected IranSeda (\texttt{book.iranseda.ir}) as our primary data source based on several critical considerations. The platform hosts over 3,800 audiobooks across many categories, ensuring broad lexical and stylistic coverage essential for TTS training. Content is produced by professional narrators in controlled recording environments at a 44.1 kHz sampling rate, providing consistent audio quality crucial for neural TTS model training. Importantly, IranSeda audiobooks are publicly accessible, making the platform a practical large-scale source for Persian audiobook data. 
	
	Although IranSeda provides audiobook recordings together with metadata, the corresponding written transcripts are not publicly available. We considered obtaining transcripts from scanned or digitised book editions using OCR, but this approach was unreliable at scale. Audiobook narrations often differ from printed editions due to abridgement, narrator edits, translation differences, or edition-specific wording, and many titles have no accessible digitised text version. Even minor text--audio mismatches would introduce systematic alignment errors and propagate noise throughout the corpus. We therefore adopted an ASR-based transcription strategy that is edition-agnostic, does not require access to source texts, and can be applied automatically at scale.
	\subsection{Intelligent Audio Segmentation}
	\label{sec:segmentation}
	Raw audiobook files typically span several hours and must be segmented into short, linguistically coherent chunks for TTS training. This is challenging because silence-based Voice Activity Detection (VAD) boundaries do not necessarily coincide with sentence boundaries: a segment may start at a natural pause but still end before the sentence is complete. We therefore use a three-phase segmentation pipeline that combines acoustic boundary detection, ASR transcription, and text-based completeness validation.
	
	\textbf{Phase 1: Acoustic Boundary Detection.}
	We apply WebRTC~\cite{webrtcvad} VAD with aggressiveness level~0 to identify silence-based candidate boundaries in each audio file; see Appendix~\ref{app:vad} for a comparison of alternative configurations.
	
\textbf{Phase 2: Transcription.}
Each candidate segment is transcribed using the Google Web Speech backend through the SpeechRecognition library~\cite{zhang2017speechrecognition}, which was selected for being free, fast, and reliable, achieving the lowest Persian ASR error rates among the tested alternatives (Appendix~\ref{app:asr_comparison}); any ASR system can be substituted. Since the API returns only the recognised text string without word-level timestamps, timestamp-based alignment is not possible, which motivates the boundary-extension procedure described below. Remaining transcription errors are removed by quality filtering in Section~\ref{sec:quality}.

	\textbf{Phase 3: Completeness Validation and Boundary Extension.}
	Each transcript is analysed by a ParsBERT-based sentence-completion classifier fine-tuned to distinguish complete from incomplete Persian sentences. Segments predicted as incomplete undergo iterative boundary extension in 0.1-second increments, up to a maximum of 5 seconds. After each extension, the segment is re-transcribed and re-evaluated by the classifier. This loop continues until the transcript satisfies the completeness criterion or the maximum extension limit is reached. Segments that remain incomplete are retained with an incompleteness flag rather than discarded, allowing downstream users to apply their own quality thresholds.
	Full details of the classifier training procedure, including the construction of complete and truncated training examples from PersianPunc~\cite{ranjbar2025persianpunc}, are provided in Appendix~\ref{app:completion_classifier}.
	\subsection{Boundary Optimization Algorithm}
	\label{sec:boundary}
\begin{figure}[t]
		\centering
		\includegraphics[width=1\linewidth]{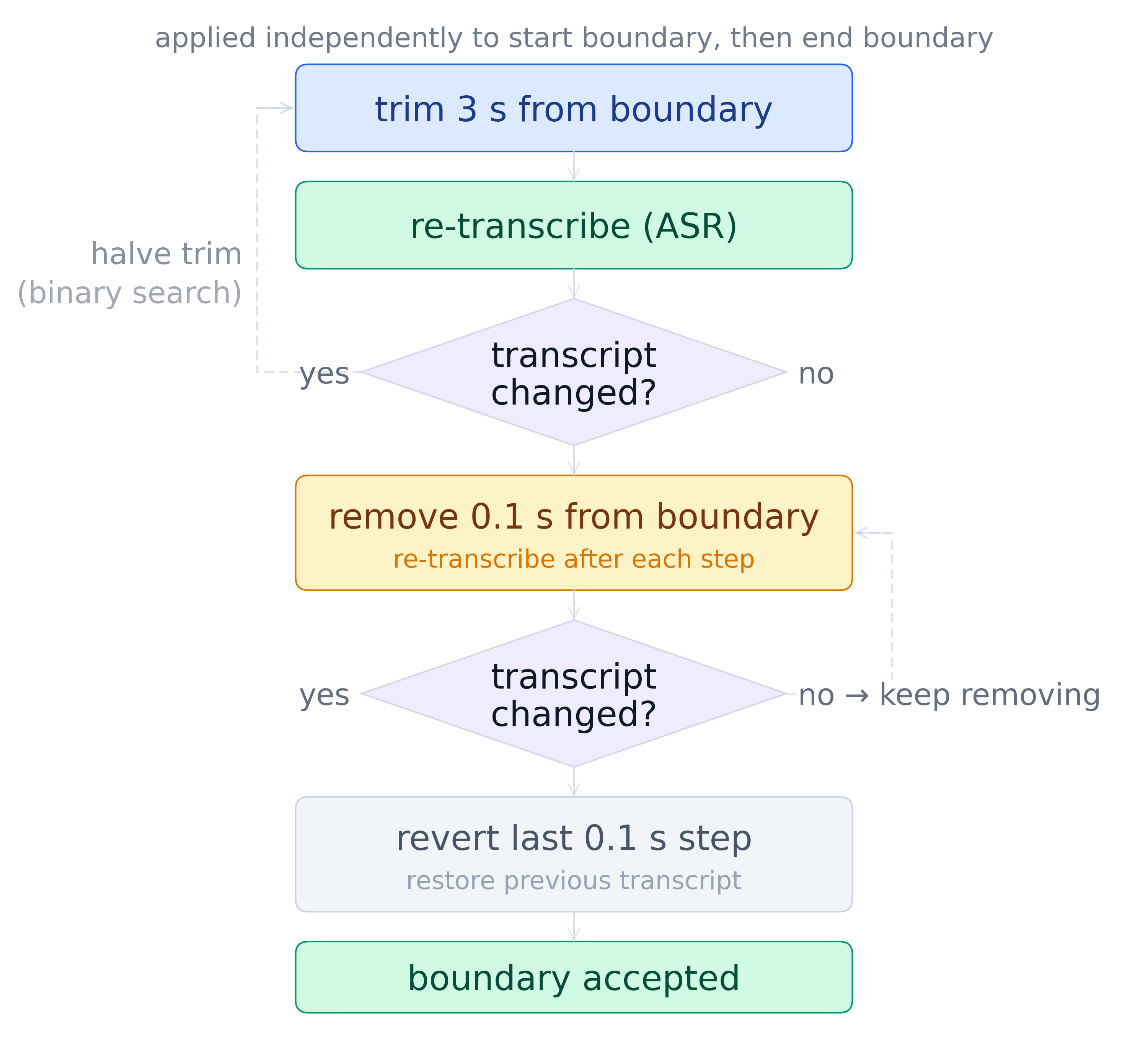}
		\caption{Boundary optimization algorithm, applied independently to the
			start and end boundaries of each segment.}
		\label{fig:boundary}
	\end{figure}

	Even with accurate transcription, audio segments may contain unwanted silence, background noise, or acoustic artifacts at boundaries that degrade TTS model performance~\cite{zen2019libritts, jung2025texttospeechsynthesiswild, nakamura2025ttsops}. Our boundary optimization algorithm employs binary search to determine optimal trimming points for both segment start and end boundaries. Figure~\ref{fig:boundary} illustrates the full algorithm.
	
	The boundary optimization is applied independently to each boundary: the start boundary is processed first, then the end boundary, using the same procedure. A hard ceiling of 3 seconds is set for the initial trimming step to prevent over-trimming in edge cases.
	
	The procedure follows a two-stage search strategy:
	
	\textbf{1. Initial Adjustment:} 3 seconds are removed from the boundary, and the trimmed audio is re-transcribed.
	
	\textbf{2. Stability Verification:} If the new transcription differs from the original in any way---that is, if even a single character changes---the trimming is deemed excessive and must be reduced. 
	
	\textbf{3. Binary Search Optimization:} The algorithm iteratively halves the trimming interval to converge on the maximum trim length that still produces an identical transcription. Binary search terminates when the transcription is character-for-character identical to the original.
	
	\textbf{4. Fine-Grained Linear Search:} After binary search converges, a linear search with 0.1-second increments is applied to achieve precise boundary alignment, continuing until the transcript changes, at which point the last step is reverted, and the boundary is accepted.
	
	In practice, the binary search converges well within the 3-second ceiling for the vast majority of segments. Leading silence is more prevalent than trailing noise in the source recordings; full trimming statistics are reported in Appendix~\ref{app:corpus_statistics} (Table~\ref{tab:trim_stats}).
	
	\subsection{Text-Audio Quality Assessment}
	\label{sec:quality}
	High-quality speech is critical for TTS training data, as low-quality samples can degrade model performance. We implement a comprehensive assessment across both audio and text dimensions. All thresholds were determined through human review of a random subset of segments: candidate threshold values were applied, and the resulting accepted and rejected samples were manually inspected until the authors judged the boundary to produce consistently clean output. The distribution of quality scores across the full dataset is provided in Appendix~\ref{app:corpus_statistics}.
	
	\subsubsection{Persian Text Quality Metrics}
	The Persian Text Quality Framework evaluates transcriptions across five dimensions: character quality, length quality, repetition score, linguistic complexity, and phonetic coverage. Each metric is normalised to $[0,1]$ and combined using fixed weights to produce a total score on $[0,1]$. Segments scoring below $0.5$ are excluded from the TTS-ready subset (Section~\ref{sec:cleaning}). Descriptive tier boundaries used for reporting, together with full metric descriptions and weights, are provided in Appendix~\ref{app:quality_details}.
		
\subsubsection{Audio Quality Metrics}
The audio quality framework assesses recordings for overall clarity, absence of distortions, and suitability for speech processing. Metrics include estimated signal-to-noise ratio, dynamic range, clipping ratio, silence ratio, background music presence (detected via inaSpeechSegmenter~\cite{ddoukhanmirex2018}), and duration. Each segment receives a composite score on a 0--100 scale; segments scoring below 75 are excluded from TTS use (Section~\ref{sec:cleaning}). Full scoring rules, metric descriptions, tier definitions, and score distributions are provided in Appendices~\ref{app:quality_details} and~\ref{app:corpus_statistics}.

\subsection{Speaker Identification}
\label{ssec:speaker_id}
Since approximately 40\% of audiobook entries lacked narrator names and some books featured multiple narrators, we apply a two-stage speaker identification pipeline based on ECAPA-TDNN embeddings~\cite{DesplanquesTD20} to assign consistent automatically inferred speaker labels across the corpus. The first stage clusters segments within each audiobook into local speaker identities; the second merges these local identities into global speaker labels across the entire corpus. Low-confidence segments and spuriously small clusters are excluded at both stages. Full implementation details, including cluster-count estimation, the clustering ensemble, outlier removal, dimensionality reduction, confidence thresholds, and cross-book merging criteria, are provided in Appendix~\ref{app:speaker_id}.

	\subsection{Final Data Cleaning and Preparation}
	\label{sec:cleaning}
	% FIX: Explained the additional filtering step that accounts for the
	% reduction from 3,007.8 h (after text quality filter) to 2,200 h (final TTS).
	% FIX: Clarified that 0.5 is the filtering threshold, distinct from the
	% descriptive tier boundary of 0.62 used in the appendix.
	We apply audio and text quality filtering to the boundary-optimised corpus: segments with audio quality scores below 75 (out of 100) and segments with text quality scores below 0.5 are removed, reducing the corpus to 2,089,749 segments (3,007.8 hours). The final TTS-ready subset is then constructed by excluding segments assigned to low-confidence speaker clusters, and segments flagged as incomplete by the sentence-completion classifier. This yields 1,364,671 segments totalling 2,199.7 hours. Since ASR transcriptions lack punctuation, we apply the ParsBERT-based Persian punctuation restoration model of~\cite{ranjbar2025persianpunc} to all segments prior to release. The release includes punctuation-restored transcripts, available metadata, automatically inferred speaker IDs, and per-segment audio and text quality scores, so users can re-filter the corpus to their own quality and speaker-confidence thresholds using the released pipeline.

	\begin{table}[t]
		\centering
		\caption{ParsVoice Dataset Statistics. We refer to the boundary-optimised corpus before TTS-specific filtering as the processed ASR-oriented subset.}
		\label{tab:parsvoice_stats}
		\resizebox{\columnwidth}{!}{%
			\begin{tabular}{|l|r|r|}
				\hline
				\textbf{Metric} & \textbf{ASR Subset} & \textbf{Final TTS Subset} \\
				\hline
				Total Hours       & 4,096.2    & 2,199.7    \\
				Segments          & 2,999,296  & 1,364,671  \\
				Total Tokens      & 29,566,314 & 17,347,614 \\
				Unique Word Forms & 309,406    & 229,860    \\
				\hline
			\end{tabular}%
		}
	\end{table}
	
	\section{ParsVoice Corpus Analysis}
	\label{sec:corpus}
	
	The current release contains 2,231 processed audiobooks, of which 1,706 are retained in the final TTS subset after quality filtering.
	
	\begin{table}[htbp]
		\centering
		\caption{ParsVoice pipeline statistics at each processing stage.}
		\label{tab:pipeline_stats}
		\resizebox{\columnwidth}{!}{%
			\begin{tabular}{|l|r|r|}
				\hline
				\textbf{Stage} & \textbf{Segments} & \textbf{Hours} \\
				\hline
				VAD output & 5,161,459 & 5,824.7 \\
				After removing empty transcripts & 3,323,679 & 5,127.3 \\
				After boundary optimization & 2,999,296 & 4,096.2 \\
				After text-audio quality filtering & 2,089,749 & 3,007.8 \\
				Final TTS subset & 1,364,671 & 2,199.7 \\
				\hline
			\end{tabular}%
		}
	\end{table}
	
	\textbf{Quality Assessment.} The mean audio-quality score is 89.5 (median 92.0), with 83.1\% of segments reaching the high-quality tier ($\geq90$) and 13.4\% in the acceptable range (75--89). Mean SNR is 41.8\,dB and 96.5\% of segments exceed 20\,dB, reflecting the controlled studio conditions of the source recordings. On the text side, the mean quality score in the final TTS subset is 0.737 (median 0.748). Combined audio and text filtering removes 30.3\% of boundary-optimised segments (Table~\ref{tab:pipeline_stats}); full score distributions and tier definitions are given in Appendices~\ref{app:quality_details} and~\ref{app:corpus_statistics}. 
	
	\textbf{Linguistic Analysis.} Following text normalisation, token and vocabulary statistics are computed using the Hazm Persian tokeniser~\cite{hazm}. Token and vocabulary counts for both subsets are reported in Table~\ref{tab:parsvoice_stats}. Segments in the final TTS subset contain an average of 12.71 words and 60.7 characters.
	
	\textbf{Transcript Verification.}
	To independently assess transcript accuracy, two native Persian speakers transcribed 500 randomly sampled segments from the TTS-ready subset. Compared with the human reference transcriptions, ParsVoice transcripts achieve 4.90\% WER and 1.81\% CER, with 69.0\% of segments transcribed perfectly. Full transcription and normalization details are provided in Appendix~\ref{app:transcript_audit}.

\textbf{Speaker Statistics.} The final TTS subset contains 1,815 automatically inferred global speaker IDs, covering both metadata-linked named narrators and audiobooks with unknown narrator names. Gender analysis was performed using narrator names present in the available metadata. Among audiobooks with available metadata, approximately 33\% of narrators are female, and 67\% are male. As a metadata-based clustering consistency check, we computed speaker-ID purity. For each global speaker ID, purity is defined as the fraction of its metadata-linked audio duration associated with the dominant narrator name. A purity of 1.0 therefore indicates that all metadata-linked audio assigned to that global ID comes from a single named narrator. Among 325 global speaker IDs with sufficient metadata-linked audio, the median purity is 1.00 and the mean is 0.871; 214 IDs (65.8\%) have purity of at least 0.90. A merge-threshold sensitivity analysis is reported in Appendix~\ref{app:speaker_validation}.

	\begin{table*}[t]
		\centering
		\caption{Evaluation results. MOS = naturalness, SMOS = speaker similarity,
			Intel. MOS = intelligibility MOS, and Spk.Sim. is ECAPA-TDNN cosine
			similarity reported as a percentage. $\dagger$ DeepMine results are taken
			directly from the original paper and were not re-evaluated by our raters.}
		\label{tab:xtts_results}
		\small
		\setlength{\tabcolsep}{8pt}
		\renewcommand{\arraystretch}{1.15}
		\begin{tabular}{@{}lcccc@{}}
			\toprule
			\textbf{System} & \textbf{MOS} & \textbf{SMOS} & \textbf{Intel. MOS} & \textbf{Spk.Sim. (\%)} \\
			\midrule
			Real Persian Speech$\dagger$
			& 4.72 & 4.86 & --- & --- \\
			
			\textbf{XTTSv2 + ParsVoice}
			& \textbf{3.60$\pm$0.09} & \textbf{4.03$\pm$0.08} & \textbf{4.03$\pm$0.08} & \textbf{80.0} \\
			
DeepMine Tacotron2 (unseen spk.)$\dagger$~\cite{Adibian2025DeepMine}
& 3.80$\pm$0.04 & --- & --- & --- \\

DeepMine FastSpeech2 (seen spk.)$\dagger$~\cite{Adibian2025DeepMine}
& 4.12$\pm$0.06 & 3.98$\pm$0.11 & --- & --- \\
			\bottomrule
		\end{tabular}
	\end{table*}
	
	\section{Evaluation: TTS Model Training}
	\label{sec:tts_eval}
	To assess whether ParsVoice is suitable for downstream TTS training, we fine-tuned XTTSv2~\cite{casanova2024xttsmassivelymultilingualzeroshot}, a state-of-the-art open-source multilingual zero-shot TTS model that supports over 17 languages and can clone a speaker's voice from a short reference clip without any fine-tuning on that speaker.
\subsection{Model Adaptation}
We adapted the GPT component of XTTSv2 to Persian while keeping the remaining modules frozen. A BPE tokeniser was trained on Persian text, and 2,500 additional Persian tokens extracted from the ParsVoice text were added to the XTTSv2 GPT vocabulary.

The model was trained on a single NVIDIA A100 GPU using AdamW with a learning rate of $5\times10^{-6}$ and weight decay of $10^{-2}$ for 340,000 steps. We used a per-device batch size of 21 with gradient accumulation over 8 steps, resulting in an effective batch size of 168. 

\subsection{Evaluation Setup}
We evaluated the fine-tuned model using 90 synthesised samples generated from random sentences and unseen Persian reference speakers drawn from the Persian Common Voice dataset. The age and gender distribution of the reference speakers is reported in Appendix~\ref{app:ref_speakers_detail}.
Subjective evaluation used 22 retained native Persian raters (of 42 participants;
retention criterion and demographics in Appendix~\ref{app:rater_details}). Each
item was rated by a median of 7 raters (mean 6.7). Raters scored naturalness using MOS, speaker similarity using SMOS, and intelligibility using an additional 1--5 intelligibility MOS. Further details about the rater pool and annotation procedure are provided in Appendix~\ref{app:rater_details}.
For objective evaluation, we measured intelligibility using WER and CER from three ASR systems: Google ASR, the transcription system used during corpus construction; a Persian-finetuned Whisper model~\cite{nezamisafa_whisper_persian_v4}, which is independent of our pipeline; and Qwen ASR 1.7B~\cite{Qwen3-ASR}. To contextualise these scores, we also evaluated the same ASR systems on the Google FLEURS Persian test set~\cite{fleurs2022arxiv}, which serves as a real-speech reference. Speaker similarity was additionally measured using ECAPA-TDNN cosine similarity between reference and synthesised speaker embeddings.

	\subsection{Subjective Evaluation Results}
	
	Table~\ref{tab:xtts_results} reports the subjective evaluation results. The fine-tuned XTTSv2 model achieves a naturalness MOS of 3.60, a speaker-similarity SMOS of 4.03, and an intelligibility MOS of 4.03 in a zero-shot setting with unseen reference speakers drawn from Persian Common Voice. For context, we also report results from \cite{Adibian2025DeepMine}, which were obtained with different raters and evaluation conditions. DeepMine FastSpeech2 achieves a MOS of 4.12 and an SMOS of 3.98 under seen-speaker evaluation and is not evaluated on unseen speakers, whereas its Tacotron2 model achieves a MOS of 3.80 under unseen-speaker evaluation. The real-speech MOS and SMOS values are also taken from \cite{Adibian2025DeepMine} as approximate upper-bound references. Because the systems differ in speaker conditions, test material, rater pools, model architectures, and input representations, these values are provided only for contextual comparison rather than as a controlled head-to-head evaluation.

	\subsection{Objective Results}
	Table~\ref{tab:wer_multiASR} reports WER and CER across three ASR systems. Because Google ASR was also used during corpus construction, we treat its results only as a diagnostic measure. The Persian-finetuned Whisper model, which is independent of the construction pipeline, provides the primary non-circular ASR-based intelligibility evaluation.
	
	Google ASR achieves 10.40\% WER and 5.92\% CER on synthesised speech, compared with 6.53\% WER and 2.55\% CER on real Persian speech from the FLEURS test set, corresponding to a WER degradation of 3.87 percentage points. The Persian-finetuned Whisper model shows a similar trend, with 17.20\% WER on synthesised speech versus 13.07\% on FLEURS, a gap of 4.13 percentage points. This consistency across an independent ASR system suggests that the relatively low error rate on synthesised speech is not solely an artefact of using Google ASR during corpus construction. Qwen ASR 1.7B produces higher absolute error rates on both real and synthesised speech, while preserving the same overall pattern.
	
	As an additional ASR-independent intelligibility check, two native Persian speakers independently transcribed the 90 synthesised samples without access to the target text. Compared with the intended sentences, the resulting human transcriptions yield a WER of 4.78\%. This provides complementary evidence that the synthesised speech is intelligible to native listeners. Further details of the human transcription protocol and agreement analysis are provided in Appendix~\ref{app:transcript_audit}.

	Finally, speaker similarity was assessed using ECAPA-TDNN~\cite{DesplanquesTD20} embeddings extracted via the SpeechBrain toolkit~\cite{speechbrain,speechbrain_v1}, computing cosine similarity between the reference and synthesised speaker representations. The resulting score of 80.0\% indicates effective zero-shot speaker adaptation across the 42 unseen reference speakers.
	
	\begin{table}[t]
		\centering
		\caption{Multi-ASR intelligibility evaluation. WER and CER are reported on 90 synthesised samples and on the Google FLEURS Persian test set as a real-speech reference. Lower WER/CER is better; the gap column shows TTS$-$Real degradation in WER.}
		\label{tab:wer_multiASR}
		\small
		\setlength{\tabcolsep}{3.5pt}
		\renewcommand{\arraystretch}{1.12}
		\begin{tabularx}{\columnwidth}{@{}Xccccc@{}}
			\toprule
			& \multicolumn{2}{c}{\textbf{TTS}} 
			& \multicolumn{2}{c}{\textbf{Real}} 
			& \textbf{Gap} \\
			\cmidrule(lr){2-3}
			\cmidrule(lr){4-5}
			\textbf{ASR System} 
			& \textbf{WER} & \textbf{CER} 
			& \textbf{WER} & \textbf{CER} 
			& \textbf{WER} \\
			\midrule
			Google ASR
			& 10.40 & 5.92 & 6.53 & 2.55 & +3.87 \\
			Whisper fine-tuned~\cite{nezamisafa_whisper_persian_v4}
			& 17.20 & 7.79 & 13.07 & 5.05 & +4.13 \\
			Qwen ASR 1.7B 
			& 34.08 & 15.32 & 22.92 & 8.56 & +11.16 \\
			\bottomrule
		\end{tabularx}
		
		\vspace{4pt}
		\par\noindent{\footnotesize All values are percentages.}
	\end{table}

	\subsection{Effect of Training-Data Scale}
	\label{sec:scale_ablation_main}
	
	To examine whether the scale of ParsVoice contributes to TTS performance, we
	fine-tuned XTTSv2 on nested 120\,h and 530\,h subsets and on the full
	2,200\,h corpus using the same tokenizer, model configuration, and evaluation
	set. Under an equal number of training epochs, increasing the amount of
	training data improves Google-ASR WER from 42.89\% for 120\,h to 20.06\% for
	530\,h and 17.83\% for the full corpus. NISQA-TTS likewise increases from
	2.968 to 3.081 and 3.165, respectively.
	
	However, equal numbers of epochs correspond to substantially different numbers
	of optimization updates across training-set sizes. We therefore also compare
	checkpoints at approximately matched optimization budgets. Under this setting,
	the differences are smaller and no training-set size is consistently best
	across all metrics, although the full-data model achieves the lowest WER and
	CER with the Persian-finetuned Whisper evaluator. The loss curves
	also indicate that the full-data model remains at an earlier stage of
	optimization, whereas the 120\,h model begins to show signs of overfitting at
	high step counts. Full numerical results and training curves are provided in
	Appendix~\ref{app:scale_ablation}.
	
	\section{Conclusion}
	\label{sec:conclusion}
	
In this work, we address the scarcity of high-quality Persian speech datasets by introducing ParsVoice, the largest publicly available Persian speech-text corpus tailored for TTS training to date. ParsVoice consists of 2,200 hours of clean, segmented speech suitable for TTS training, covering 1,364,671 segments from 1,815 automatically inferred speaker IDs. In addition, the broader processed corpus contains 4,096.2 hours of audio, making it useful for a wider range of Persian speech research tasks.

Alongside the dataset, we provide a scalable and automated pipeline for dataset creation. This pipeline incorporates several key parts, including a BERT-based model to ensure sentence completeness, a binary search algorithm for precise audio-text boundary optimization, and comprehensive audio-text quality assessment frameworks specifically designed for Persian. Together, the dataset and the pipeline provide a valuable resource for advancing Persian speech research and TTS development.

We validated ParsVoice by fine-tuning XTTSv2, achieving a naturalness MOS of 3.60/5 and a speaker similarity SMOS of 4.03/5 (Section~\ref{sec:tts_eval}). Multi-ASR intelligibility evaluation using both the transcription-pipeline ASR and a Persian-finetuned Whisper model provides complementary evidence that the synthesised speech remains intelligible under evaluation. These results support the usability of ParsVoice for training multi-speaker Persian TTS systems, addressing a critical resource gap in Persian language technology.

\section{Limitations}
\label{sec:limitations}
Despite the scale and quality of ParsVoice, several limitations should be acknowledged. First, the corpus is derived entirely from audiobooks, which means the speaking style is predominantly formal and narrative. Spontaneous conversational speech, which differs substantially in prosody, pace, and vocabulary, is not represented. TTS models trained on ParsVoice may therefore produce speech that sounds natural in read-aloud contexts but less so in conversational applications. Second, our transcription pipeline relies on a proprietary ASR system without access to the original book texts, which introduces a small but non-zero transcription error rate. While our multi-stage quality filtering mitigates this, some residual errors may remain in the dataset. Third, the evaluation of ParsVoice was conducted by fine-tuning XTTSv2, a single model architecture. Due to limited access to GPU compute, we were unable to train multiple model architectures or conduct extensive hyperparameter searches, which limits the generalisability of the validation results. Similarly, we were unable to evaluate against restricted-access baseline models such as DeepMine Multi-TTS, meaning our MOS comparison relies on scores reported in the original papers rather than a controlled head-to-head evaluation with shared raters. Finally, around 40\% of audiobooks lack narrator metadata (Section~\ref{sec:corpus}), with two consequences. The metadata-linked portion of the corpus is gender-imbalanced, which may affect the quality of TTS voices synthesised for female speakers relative to male speakers, and the full-corpus gender distribution remains only partially observed. Speaker identity for the unlabelled recordings is also assigned through automated clustering rather than verified ground truth; although the ECAPA-TDNN-based pipeline provides useful global speaker IDs, these labels should be interpreted as automatically inferred speaker identities rather than manually verified speaker annotations. Segments whose speaker identity is confirmed by narrator metadata are flagged in the release, allowing users who require verified speaker labels to restrict to that subset.

	\section{Ethical Considerations}
	\label{sec:ethics}
	ParsVoice is derived from publicly accessible audiobooks on IranSeda. We do not redistribute complete recordings, only short quality-filtered excerpts, consistent with Article 7 of Iran's Copyright Act, which permits quotation from published works for scientific and educational purposes with attribution. Copyright in the underlying recordings remains with IranSeda and the narrators; we make no ownership claim. The pipeline code is released under Apache 2.0 and our annotations (transcripts, speaker IDs, quality scores) under CC BY-NC 4.0. Audio segments are distributed for non-commercial research only, under gated access on Hugging Face. Rights holders may request removal; affected audio will be excluded from future releases.
	
	Because ParsVoice may support voice-cloning-capable TTS systems, potential misuse includes impersonation, fraud, deceptive synthetic media, and disinformation. Such uses are outside the intended scope of the dataset. We encourage downstream users to obtain appropriate speaker consent for deployed cloned voices and to use provenance or watermarking mechanisms for publicly released synthetic speech.

	\bibliography{custom}

\begin{thebibliography}{31}
\providecommand{\natexlab}[1]{#1}

\bibitem[{Adibian et~al.(2025)Adibian, Zeinali, and
  Barmaki}]{Adibian2025DeepMine}
Majid Adibian, Hossein Zeinali, and Soroush Barmaki. 2025.
\newblock \href {https://doi.org/10.1007/s10579-025-09807-6}
  {Deepmine-multi-tts: a persian speech corpus for multi-speaker
  text-to-speech}.
\newblock \emph{Language Resources and Evaluation}, 59:2245--2264.

\bibitem[{Ardila et~al.(2020)Ardila, Branson, Davis, Kohler, Meyer, Henretty,
  Morais, Saunders, Tyers, and Weber}]{ardila2020common}
Rosana Ardila, Megan Branson, Kelly Davis, Michael Kohler, Josh Meyer, Michael
  Henretty, Reuben Morais, Lindsay Saunders, Francis Tyers, and Gregor Weber.
  2020.
\newblock \href {https://aclanthology.org/2020.lrec-1.520/} {Common voice: A
  massively-multilingual speech corpus}.
\newblock In \emph{Proceedings of the Twelfth Language Resources and Evaluation
  Conference}, pages 4218--4222, Marseille, France. European Language Resources
  Association.

\bibitem[{Casanova et~al.(2024)Casanova, Davis, Gölge, Göknar, Gulea, Hart,
  Aljafari, Meyer, Morais, Olayemi, and
  Weber}]{casanova2024xttsmassivelymultilingualzeroshot}
Edresson Casanova, Kelly Davis, Eren Gölge, Görkem Göknar, Iulian Gulea,
  Logan Hart, Aya Aljafari, Joshua Meyer, Reuben Morais, Samuel Olayemi, and
  Julian Weber. 2024.
\newblock \href {https://doi.org/10.21437/Interspeech.2024-2016} {{XTTS: a
  Massively Multilingual Zero-Shot Text-to-Speech Model}}.
\newblock In \emph{{Interspeech 2024}}, pages 4978--4982.

\bibitem[{Conneau et~al.(2023)Conneau, Ma, Khanuja, Zhang, Axelrod, Dalmia,
  Riesa, Rivera, and Bapna}]{fleurs2022arxiv}
Alexis Conneau, Min Ma, Simran Khanuja, Yu~Zhang, Vera Axelrod, Siddharth
  Dalmia, Jason Riesa, Clara Rivera, and Ankur Bapna. 2023.
\newblock \href {https://doi.org/10.1109/SLT54892.2023.10023141} {Fleurs:
  Few-shot learning evaluation of universal representations of speech}.
\newblock In \emph{2022 IEEE Spoken Language Technology Workshop (SLT)}, pages
  798--805.

\bibitem[{Desplanques et~al.(2020)Desplanques, Thienpondt, and
  Demuynck}]{DesplanquesTD20}
Brecht Desplanques, Jenthe Thienpondt, and Kris Demuynck. 2020.
\newblock {ECAPA-TDNN: Emphasized Channel Attention, propagation and
  aggregation in TDNN based speaker verification}.
\newblock In \emph{Interspeech 2020}, pages 3830--3834.

\bibitem[{Doukhan et~al.(2018)Doukhan, Lechapt, Evrard, and
  Carrive}]{ddoukhanmirex2018}
David Doukhan, Eliott Lechapt, Marc Evrard, and Jean Carrive. 2018.
\newblock Ina’s mirex 2018 music and speech detection system.
\newblock In \emph{Music Information Retrieval Evaluation eXchange (MIREX
  2018)}.

\bibitem[{Farahani et~al.(2021)Farahani, Gharachorloo, Farahani, and
  Manthouri}]{Farahani_2021}
Mehrdad Farahani, Mohammad Gharachorloo, Marzieh Farahani, and Mohammad
  Manthouri. 2021.
\newblock \href {https://doi.org/10.1007/s11063-021-10528-4} {Parsbert:
  Transformer-based model for persian language understanding}.
\newblock \emph{Neural Processing Letters}, 53(6):3831–3847.

\bibitem[{Ito and Johnson(2017)}]{ito2017lj}
Keith Ito and Linda Johnson. 2017.
\newblock The lj speech dataset.
\newblock \url{https://keithito.com/LJ-Speech-Dataset/}.

\bibitem[{Jung et~al.(2024)Jung, Zhang, Maiti, Wu, Wang, Kim, Matsunaga, Um,
  Tian, Shim et~al.}]{jung2025texttospeechsynthesiswild}
Jee-weon Jung, Wangyou Zhang, Soumi Maiti, Yihan Wu, Xin Wang, Ji-Hoon Kim,
  Yuta Matsunaga, Seyun Um, Jinchuan Tian, Hye-jin Shim, and 1 others. 2024.
\newblock Text-to-speech synthesis in the wild.
\newblock \emph{arXiv preprint arXiv:2409.08711}.

\bibitem[{Kalahroodi et~al.(2026)Kalahroodi, Faili, and
  Shakery}]{ranjbar2025persianpunc}
Mohammad Javad~Ranjbar Kalahroodi, Heshaam Faili, and Azadeh Shakery. 2026.
\newblock \href {https://aclanthology.org/2026.silkroadnlp-1.11/}
  {{P}ersian{P}unc: A large-scale dataset and {BERT}-based approach for
  {P}ersian punctuation restoration}.
\newblock In \emph{The Proceedings of the First Workshop on {NLP} and {LLM}s
  for the {I}ranian Language Family}, pages 105--113, Rabat, Morocco.
  Association for Computational Linguistics.

\bibitem[{McAuliffe et~al.(2017)McAuliffe, Socolof, Mihuc, Wagner, and
  Sonderegger}]{mcauliffe17_interspeech}
Michael McAuliffe, Michaela Socolof, Sarah Mihuc, Michael Wagner, and Morgan
  Sonderegger. 2017.
\newblock \href {https://doi.org/10.21437/Interspeech.2017-1386} {{Montreal
  Forced Aligner: Trainable Text-Speech Alignment Using Kaldi}}.
\newblock In \emph{{Interspeech 2017}}, pages 498--502.

\bibitem[{Mittag and Möller(2020)}]{mittag20_interspeech}
Gabriel Mittag and Sebastian Möller. 2020.
\newblock \href {https://doi.org/10.21437/Interspeech.2020-2382} {{Deep
  Learning Based Assessment of Synthetic Speech Naturalness}}.
\newblock In \emph{{Interspeech 2020}}, pages 1748--1752.

\bibitem[{Nezamisafa(2025)}]{nezamisafa_whisper_persian_v4}
Nezamisafa. 2025.
\newblock whisper-persian-v4.
\newblock \url{https://huggingface.co/nezamisafa/whisper-persian-v4}.
\newblock Fine-tuned version of OpenAI Whisper-large-v3 for Persian automatic
  speech recognition. Accessed: 2026-05-25.

\bibitem[{Panayotov et~al.(2015)Panayotov, Chen, Povey, and
  Khudanpur}]{panayotov2015librispeech}
Vassil Panayotov, Guoguo Chen, Daniel Povey, and Sanjeev Khudanpur. 2015.
\newblock \href {https://doi.org/10.1109/ICASSP.2015.7178964} {Librispeech: An
  asr corpus based on public domain audio books}.
\newblock In \emph{2015 IEEE International Conference on Acoustics, Speech and
  Signal Processing (ICASSP)}, pages 5206--5210.

\bibitem[{Parcollet et~al.(2022)Parcollet, Ravanelli, Plantinga, Rouhe,
  Cornell, Lugosch, Subakan, Dawalatabad, Heba, Zhong, Chou, Yeh, Fu, Liao,
  Rastorgueva, Grondin, Aris, Na, Gao, de~Mori, and Bengio}]{speechbrain}
Titouan Parcollet, Mirco Ravanelli, Peter Plantinga, Aku Rouhe, Samuele
  Cornell, Loren Lugosch, Cem Subakan, Nauman Dawalatabad, Abdelwahab Heba,
  Jianyuan Zhong, Ju-Chieh Chou, Sung-Lin Yeh, Szu-Wei Fu, Chien-Feng Liao,
  Elena Rastorgueva, Fran{\c c}ois Grondin, William Aris, Hwidong Na, Yan Gao,
  and 2 others. 2022.
\newblock \href {https://hal.science/hal-03601303} {{SpeechBrain: A
  General-Purpose Speech Toolkit}}.
\newblock Preprint.

\bibitem[{Pratap et~al.(2020)Pratap, Xu, Sriram, Synnaeve, and
  Collobert}]{pratap2020mls}
Vineel Pratap, Qiantong Xu, Anuroop Sriram, Gabriel Synnaeve, and Ronan
  Collobert. 2020.
\newblock \href {https://doi.org/10.21437/interspeech.2020-2826} {Mls: A
  large-scale multilingual dataset for speech research}.
\newblock In \emph{Interspeech 2020}, interspeech\_2020, page 2757–2761.
  ISCA.

\bibitem[{Qharabagh et~al.(2024)Qharabagh, Dehghanian, and
  Rabiee}]{fetrat2024mana}
Mahta~Fetrat Qharabagh, Zahra Dehghanian, and Hamid~R. Rabiee. 2024.
\newblock \href {https://arxiv.org/abs/2409.07259} {Manatts persian: a recipe
  for creating tts datasets for lower resource languages}.
\newblock \emph{Preprint}, arXiv:2409.07259.

\bibitem[{Radford et~al.(2023)Radford, Kim, Xu, Brockman, Mcleavey, and
  Sutskever}]{radford2022whisper}
Alec Radford, Jong~Wook Kim, Tao Xu, Greg Brockman, Christine Mcleavey, and
  Ilya Sutskever. 2023.
\newblock \href {https://proceedings.mlr.press/v202/radford23a.html} {Robust
  speech recognition via large-scale weak supervision}.
\newblock In \emph{Proceedings of the 40th International Conference on Machine
  Learning}, volume 202 of \emph{Proceedings of Machine Learning Research},
  pages 28492--28518. PMLR.

\bibitem[{Ravanelli et~al.(2024)Ravanelli, Parcollet, Moumen, \{de Langen\},
  Subakan, Plantinga, Wang, Mousavi, Libera, Ploujnikov, Paissan, Borra, Zaiem,
  Zhao, Zhang, Karakasidis, Yeh, Champion, Rouhe, Braun, Mai, Zuluaga-Gomez,
  Mousavi, Nautsch, Nguyen, Liu, Sagar, Duret, Mdhaffar, Laperri{\`e}re,
  Rouvier, \{De Mori\}, and Est{\`e}ve}]{speechbrain_v1}
Mirco Ravanelli, Titouan Parcollet, Adel Moumen, Sylvain \{de Langen\}, Cem
  Subakan, Peter Plantinga, Yingzhi Wang, Pooneh Mousavi, \{Luca Della\}
  Libera, Artem Ploujnikov, Francesco Paissan, Davide Borra, Salah Zaiem, Zeyu
  Zhao, Shucong Zhang, Georgios Karakasidis, \{Sung Lin\} Yeh, Pierre Champion,
  Aku Rouhe, and 14 others. 2024.
\newblock Open-source conversational ai with speechbrain 1.0.
\newblock \emph{Journal of Machine Learning Research}, 25.
\newblock Publisher Copyright: {\textcopyright} 2024 Mirco Ravanelli, Titouan
  Parcollet, et al.

\bibitem[{Roshan(2014)}]{hazm}
Roshan. 2014.
\newblock \href {https://github.com/roshan-research/hazm} {hazm: Python library
  for digesting persian text}.
\newblock Available at \url{https://github.com/roshan-research/hazm}.

\bibitem[{Seki et~al.(2025)Seki, Takamichi, Saeki, and
  Saruwatari}]{nakamura2025ttsops}
Kentaro Seki, Shinnosuke Takamichi, Takaaki Saeki, and Hiroshi Saruwatari.
  2025.
\newblock \href {https://doi.org/10.1109/TASLPRO.2025.3633089} {Ttsops: A
  closed-loop corpus optimization framework for training multi-speaker tts
  models from dark data}.
\newblock \emph{IEEE Transactions on Audio, Speech and Language Processing},
  33:4956--4970.

\bibitem[{Shamgholi et~al.(2023)Shamgholi, Saeedi, Peymanfard, Alhabib, and
  Zeinali}]{shamgholi2023armantts}
Mohammd~Hasan Shamgholi, Vahid Saeedi, Javad Peymanfard, Leila Alhabib, and
  Hossein Zeinali. 2023.
\newblock \href {https://arxiv.org/abs/2304.03585} {Armantts single-speaker
  persian dataset}.
\newblock \emph{Preprint}, arXiv:2304.03585.

\bibitem[{Shi et~al.(2026)Shi, Wang, Guo, Wang, Zhang, Zhang, Guo, Hao, Xi,
  Yang, Xu, Zhou, and Lin}]{Qwen3-ASR}
Xian Shi, Xiong Wang, Zhifang Guo, Yongqi Wang, Pei Zhang, Xinyu Zhang, Zishan
  Guo, Hongkun Hao, Yu~Xi, Baosong Yang, Jin Xu, Jingren Zhou, and Junyang Lin.
  2026.
\newblock Qwen3-asr technical report.
\newblock \emph{arXiv preprint arXiv:2601.21337}.

\bibitem[{SileroTeam(2024)}]{silero_vad}
SileroTeam. 2024.
\newblock Silero vad: pre-trained enterprise-grade voice activity detector
  (vad), number detector and language classifier.
\newblock \url{https://github.com/snakers4/silero-vad}.

\bibitem[{Vaswani et~al.(2017)Vaswani, Shazeer, Parmar, Uszkoreit, Jones,
  Gomez, Kaiser, and Polosukhin}]{vaswani2017attention}
Ashish Vaswani, Noam Shazeer, Niki Parmar, Jakob Uszkoreit, Llion Jones,
  Aidan~N Gomez, \L~ukasz Kaiser, and Illia Polosukhin. 2017.
\newblock \href
  {https://proceedings.neurips.cc/paper_files/paper/2017/file/3f5ee243547dee91fbd053c1c4a845aa-Paper.pdf}
  {Attention is all you need}.
\newblock In \emph{Advances in Neural Information Processing Systems},
  volume~30. Curran Associates, Inc.

\bibitem[{Wang et~al.(2021)Wang, Riviere, Lee, Wu, Talnikar, Haziza,
  Williamson, Pino, and Dupoux}]{wang2021voxpopuli}
Changhan Wang, Morgane Riviere, Ann Lee, Anne Wu, Chaitanya Talnikar, Daniel
  Haziza, Mary Williamson, Juan Pino, and Emmanuel Dupoux. 2021.
\newblock \href {https://doi.org/10.18653/v1/2021.acl-long.80} {{V}ox{P}opuli:
  A large-scale multilingual speech corpus for representation learning,
  semi-supervised learning and interpretation}.
\newblock In \emph{Proceedings of the 59th Annual Meeting of the Association
  for Computational Linguistics and the 11th International Joint Conference on
  Natural Language Processing (Volume 1: Long Papers)}, pages 993--1003,
  Online. Association for Computational Linguistics.

\bibitem[{Wiseman()}]{webrtcvad}
Wiseman.
\newblock Webrtc voice activity detector.
\newblock \url{https://github.com/wiseman/py-webrtcvad}.
\newblock Accessed: 2025-09-14.

\bibitem[{Yamagishi et~al.(2019)Yamagishi, Veaux, and
  MacDonald}]{veaux2017cstr}
Junichi Yamagishi, Christophe Veaux, and Kirsten MacDonald. 2019.
\newblock \href {https://doi.org/10.7488/ds/2645} {Cstr vctk corpus: English
  multi-speaker corpus for cstr voice cloning toolkit (version 0.92)}.

\bibitem[{Zeinali et~al.(2019)Zeinali, Burget, and
  Černocký}]{zeinali2019deepmine}
Hossein Zeinali, Lukáš Burget, and Jan~"Honza'' Černocký. 2019.
\newblock \href {https://arxiv.org/abs/1912.03627} {A multi purpose and large
  scale speech corpus in persian and english for speaker and speech
  recognition: the deepmine database}.
\newblock \emph{Preprint}, arXiv:1912.03627.

\bibitem[{Zen et~al.(2019)Zen, Dang, Clark, Zhang, Weiss, Jia, Chen, and
  Wu}]{zen2019libritts}
Heiga Zen, Viet Dang, Rob Clark, Yu~Zhang, Ron~J. Weiss, Ye~Jia, Zhifeng Chen,
  and Yonghui Wu. 2019.
\newblock \href {https://doi.org/10.21437/Interspeech.2019-2441} {{LibriTTS: A
  Corpus Derived from LibriSpeech for Text-to-Speech}}.
\newblock In \emph{{Interspeech 2019}}, pages 1526--1530.

\bibitem[{Zhang(2017)}]{zhang2017speechrecognition}
Anthony Zhang. 2017.
\newblock Speech recognition (version 3.11).
\newblock Software. Available from
  \url{https://github.com/Uberi/speech_recognition}.
\newblock Accessed: 2025.

\end{thebibliography}
	
\newpage
\pagebreak

	\appendix

	% ─────────────────────────────────────────────────────────────────────────────
	\section{VAD Method Comparison}
	\label{app:vad}
	To justify the choice of the VAD algorithm used in Phase~1 of the segmentation
	pipeline (Section~\ref{sec:segmentation}), we evaluated six VAD configurations
	on a random sample of ten audiobooks. The evaluation metric is \emph{completion
		rate}: the percentage of produced segments that the ParsBERT sentence-completion
	classifier identifies as complete sentences.
	\begin{table}[h]
		\centering
		\small
		\begin{tabular}{lc}
			\toprule
			\textbf{VAD Method} & \textbf{Completion Rate (\%)} \\
			\midrule
			\textbf{WebRTC Level 0} & \textbf{75.31} \\
			WebRTC Level 1          & 73.43 \\
			WebRTC Level 2          & 59.62 \\
			WebRTC Level 3          & 31.60 \\
			Silero ~\cite{silero_vad}            & 42.02 \\
			Whisper-based VAD       & 52.10 \\
			\bottomrule
		\end{tabular}
		\caption{Completion rates of segments produced by different VAD methods on
			ten randomly selected audiobooks.}
		\label{tab:vad_comparison}
	\end{table}
	WebRTC Level~0 achieved the highest completion rate while remaining lightweight and scalable; Silero and Whisper-based VAD are additionally heavier to run. We therefore selected WebRTC Level~0 as the segmentation backend.
	% ─────────────────────────────────────────────────────────────────────────────
	\section{ASR System Comparison}
	\label{app:asr_comparison}
	We compared three ASR systems on 1,000 randomly selected samples from Persian
	Common Voice~\cite{ardila2020common} to select a transcription backend.
	\begin{table}[h]
		\centering
		\small
		\setlength{\tabcolsep}{4pt}
		\begin{tabularx}{\linewidth}{Xcc}
			\toprule
			\textbf{ASR System} & \textbf{WER} & \textbf{CER} \\
			\midrule
			\textbf{Google Speech Rec.} & \textbf{20.10} & \textbf{7.03} \\
			Whisper-large-v3 Persian-finetuned~\cite{nezamisafa_whisper_persian_v4} & 32.50 & 11.14 \\
			Whisper-large-v3~\cite{radford2022whisper} & 43.77 & 15.62 \\
			\bottomrule
		\end{tabularx}
		\caption{Persian ASR comparison on 1,000 Persian Common Voice samples. Lower is better.}
		\label{tab:asr_comparison}
	\end{table}
	
	Google Speech Recognition achieves the lowest WER (20.10\%) and CER (7.03\%)
	and was therefore selected as the transcription backend.
	% ─────────────────────────────────────────────────────────────────────────────
	\section{Sentence Completion Classifier}
	\label{app:completion_classifier}
	The classifier is fine-tuned from ParsBERT~\cite{Farahani_2021} on synthetic
	data from PersianPunc~\cite{ranjbar2025persianpunc}. Positive examples are
	complete sentences; negatives are generated by removing up to five words from
	the sentence end or truncating the final word by up to five characters,
	reflecting the tendency of VAD boundaries to cut sentences at their endings.
	Incomplete examples are oversampled at a 3:2 ratio. Training runs for 3 epochs
	with Adam (lr $=2\times10^{-5}$, batch size 128, weight decay 0.01). On a
	10,000-sample held-out set the classifier achieves 97.4\% F1 across both
	classes.
	% ─────────────────────────────────────────────────────────────────────────────
	\section{Audio and Text Quality Scoring Details}
	\label{app:quality_details}
	\subsection{Audio Quality Scoring}
	Each recording receives a composite score on a 0--100 scale. Scores of
	$\geq90$ are \emph{high quality}; 75--89 are \emph{acceptable}; below 75 are
	\emph{low quality}. Scoring rules are listed in
	Table~\ref{tab:audio_scoring}.
	\begin{table}[h]
		\centering
		\caption{Audio quality scoring rules. Adjustments are summed and clamped
			to $[0,100]$.}
		\label{tab:audio_scoring}
		\small
		\setlength{\tabcolsep}{5pt}
		\begin{tabular}{p{3.8cm}lc}
			\toprule
			\textbf{Metric} & \textbf{Condition} & \textbf{Adj.} \\
			\midrule
			\textbf{SNR}              & $>20$\,dB  & $+35$ \\
			& $10$--$20$\,dB & $+25$ \\
			& $5$--$10$\,dB  & $+15$ \\
			\midrule
			\textbf{Dynamic range}    & $>20$\,dB  & $+15$ \\
			& $15$--$20$\,dB & $+12$ \\
			& $10$--$15$\,dB & $+10$ \\
			\midrule
			\textbf{Duration}         & $3$--$15$\,s & $+15$ \\
			& $15$--$30$\,s & $+5$  \\
			\midrule
			\textbf{No background music} & clean (binary) & $+35$ \\
			\midrule
			\textbf{Clipping}         & $>5\%$     & $-20$ \\
			& $1$--$5\%$ & $-10$ \\
			& $0.1$--$1\%$ & $-5$ \\
			\midrule
			\textbf{Silence}          & $>50\%$    & $-15$ \\
			& $30$--$50\%$ & $-10$ \\
			& $20$--$30\%$ & $-5$  \\
			\bottomrule
		\end{tabular}
	\end{table}
	\subsection{Text Quality Scoring}
	Each transcription is scored on a 0--1 scale across five weighted dimensions.
	(Table~\ref{tab:text_scoring}). 
	Tier thresholds for descriptive reporting: $\geq0.78$ = \emph{high quality}; $0.62$--$0.78$ =
	\emph{mid quality}; $<0.62$ = \emph{low quality}. Note that these tiers are used only for
	analysis and reporting; the actual filtering threshold that excludes segments from the TTS-ready
	corpus is $0.5$ (Section~\ref{sec:cleaning}).
	\begin{table}[t]
		\centering
		\caption{Text quality dimensions, weights, and descriptions.}
		\label{tab:text_scoring}
		\small
		\setlength{\tabcolsep}{3pt}
		\renewcommand{\arraystretch}{1.15}
		\begin{tabularx}{\columnwidth}{p{2.2cm} c X}
			\toprule
			\textbf{Dimension} & \textbf{Weight} & \textbf{Description} \\
			\midrule
			Character quality   & 0.25  & Fraction of valid Persian characters or digits; penalises Arabic-script substitutes. \\
			Length quality      & 0.20  & Word and character counts in an acceptable range (3--30 words, 20--200 characters). \\
			Repetition score    & 0.20  & Lexical diversity; penalises excessive word repetition. \\
			Linguistic complexity & 0.175 & Combines the common-word ratio and average word length. \\
			Phonetic coverage   & 0.175 & Diversity of Persian characters; favours broad phoneme coverage. \\
			\bottomrule
		\end{tabularx}
	\end{table}
	% ─────────────────────────────────────────────────────────────────────────────
	\section{Corpus Statistics}
	\label{app:corpus_statistics}
	\subsection{Pipeline Stage Statistics}
	Segment counts and hours at each processing stage are reported in
	Table~\ref{tab:pipeline_stats} in Section~\ref{sec:corpus}. After
	transcription and before boundary optimization, segments have a mean duration
	of 5.55\,s (median 4.20\,s). Only 1.1\% of segments are shorter than 1\,s
	and 1.9\% exceed 20\,s, confirming that VAD-based segmentation produces
	broadly TTS-suitable lengths even before further refinement. Of the 3,323,679
	segments with non-empty transcripts, 86.3\% were classified as complete by the
	ParsBERT classifier; the remaining 13.7\% underwent iterative boundary
	extension with a mean of 17.7 extension steps.
	\subsection{Boundary optimization}
	% FIX: Corrected the segment count description. , the text stated the
	% algorithm "processed 2,999,296 post-transcription segments", which confused
	% the output count with the input count. The algorithm received 3,323,679
	% non-empty transcript segments as input and retained 2,999,296 after
	% discarding transcription-unstable segments.
	The boundary optimization algorithm processed 3,323,679 non-empty transcript segments. Of these, 324,383 segments were discarded due to transcription instability or failed re-transcription during the boundary search, accounting for approximately 524 hours of audio. The remaining 2,999,296 segments were retained and trimmed. Table~\ref{tab:trim_stats} summarises trimming statistics for the retained segments.
	\begin{table}[t]
		\centering
		\caption{Boundary optimization trimming statistics across 2,999,296
			retained segments.}
		\label{tab:trim_stats}
		\small
		\setlength{\tabcolsep}{4pt}
		\resizebox{\linewidth}{!}{%
			\begin{tabular}{lrr}
				\toprule
				\textbf{Metric} & \textbf{Start} & \textbf{End} \\
				\midrule
				Segments trimmed          & 2,434,154 (81.2\%) & 1,510,777 (50.4\%) \\
				Mean trim, all segs (ms)  & 426.9              & 181.6              \\
				Mean trim, trimmed (ms)   & 526.0              & 360.5              \\
				\midrule
				% FIX: Clarified the 11.0% denominator. The percentage is computed
				% relative to the 4,603.2 h pre-trim duration of the 2,999,296
				% retained segments, not relative to the 5,127.3 h post-transcription
				% total (which also includes discarded segments).
				\multicolumn{3}{l}{Total audio removed by trimming: 507.0\,h (11.0\% of pre-trim duration of retained segments, 4,603.2\,h)} \\
				\bottomrule
			\end{tabular}%
		}
	\end{table}
\subsection{Audio Quality}
Summary audio-quality statistics are reported in Section~\ref{sec:corpus}. Of all scored segments, 3.5\% fall below the exclusion threshold of 75 and are removed. Background music was detected in 4.3\% of segments by inaSpeechSegmenter~\cite{ddoukhanmirex2018}.

\subsection{Text Quality}
Summary text-quality statistics are reported in Section~\ref{sec:corpus}. Sub-metric means in the final TTS subset are: character quality 0.851, length quality 0.900, repetition score 0.597, linguistic complexity 0.586, and phonetic coverage 0.699.
	% ─────────────────────────────────────────────────────────────────────────────
	\section{Speaker Identification Details}
	\label{app:speaker_id}
	\paragraph{Local clustering.}
	Segment-level ECAPA-TDNN embeddings are preprocessed by removing statistical
	outliers (z-score threshold 3.0), L2-normalised, and PCA-reduced when the
	embedding dimension exceeds 512. The number of speakers $k^*$ per book is
	estimated by consensus among the Silhouette score, the gap statistic, and
	DBSCAN-based estimates. Clustering uses an ensemble of Agglomerative (cosine
	distance, average linkage), Spectral, and Gaussian Mixture Model methods, with
	the result selected by the highest silhouette score. Each segment receives a
	confidence score $c_i = 0.6\,\tilde{s}_i + 0.4\,d_i$, where $\tilde{s}_i$ is
	the normalised silhouette score and $d_i$ is the distance-based confidence.
	Segments with $c_i < 0.4$ and clusters smaller than $1/k^*$ of the
	high-confidence pool are excluded; speaker groups with fewer than 10\% of
	the book's high-confidence segments are discarded entirely, preventing
	fragmented utterances from entering the final subset.
	\paragraph{Global merging.}
	Local cluster centroids are compared across all books using pairwise cosine similarity. Agglomerative clustering with average linkage and a similarity threshold of 0.90 forms global speaker-cluster IDs. Cross-book merges are accepted only when the matched speakers share at least 100 segments in total; otherwise they are kept as separate single-book identities. 

\subsection{Speaker-ID Validation}
\label{app:speaker_validation}

Narrator metadata covers 83.2\% of segments. For each global speaker ID, we define purity as the fraction of its metadata-linked audio associated with the dominant narrator name. Across 325 IDs with sufficient metadata, median purity is 1.000 and mean purity is 0.871. Overall, 10.8\% of IDs are associated with more than one narrator name. Because narrator metadata are provided at the book level, genuinely multi-narrator books may also create such conflicts; this rate therefore provides an upper bound on speaker mixing.

Table~\ref{tab:thr_sweep} shows sensitivity to the global merge threshold. Higher thresholds make cross-book merging more conservative and produce more speaker IDs. Purity remains stable through 0.90 but decreases at 0.95, indicating that an overly strict threshold does not improve metadata consistency. We therefore use 0.90. Here, $N_2$ denotes the effective number of speakers after accounting for highly unequal cluster sizes. Because the resulting IDs are automatically inferred and may fragment the same speaker across multiple clusters, the reported 1,815 IDs should not be interpreted as a verified count of distinct speakers.

\begin{table}[h] \centering \small \caption{Global-merge threshold sensitivity.} \label{tab:thr_sweep} \begin{tabular}{lrrr} \toprule \textbf{Threshold} & \textbf{IDs} & \textbf{$N_2$} & \textbf{Median purity} \\ \midrule 0.75 & 921 & 64.3 & 1.000 \\ 0.80 & 1165 & 94.5 & 1.000 \\ 0.85 & 1472 & 98.2 & 1.000 \\ 0.90 & 1815 & 142.1 & 1.000 \\ 0.95 & 2448 & 466.2 & 0.832 \\ \bottomrule \end{tabular} \end{table}

	\section{Evaluation Details}
	
	\subsection{Annotation Interface and Procedure}
	\label{app:annotation_interface}
	
	Ratings were collected via \textit{Goyar}, a custom Telegram-based annotation bot developed for this study. Raters accessed the interface through their personal Telegram accounts at their own convenience. Before the session began, each rater received the following standardised instructions: complete the evaluation in a quiet environment, and use in-ear headphones (earphones) throughout, to ensure consistent and distraction-free listening conditions.
	
	Each session comprised 90 evaluation items presented sequentially. For each item, the bot displayed (i) the item index out of 90, (ii) the Persian reference text, (iii) a reference audio clip recorded by the target speaker, and (iv) the corresponding synthesised audio clip. Raters were explicitly instructed to listen to the \emph{reference audio} first in order to familiarise themselves with the target speaker's voice characteristics before listening to the \emph{synthesised audio}. Ratings were submitted directly via the bot's inline response buttons.
	
	Each synthesised sample was scored on three criteria using the following scales:
	
	\paragraph{1. Naturalness (MOS).}
	How natural and human-like does the synthesised voice sound?
	Raters were asked to consider whether the voice sounded robotic, whether prosody and intonation were unusual, and whether pauses and pronunciation were natural.
	\begin{itemize}
		\item 1 = Completely unnatural, machine-like
		\item 2 = Somewhat unnatural with clear defects
		\item 3 = Acceptable but not fully natural
		\item 4 = Almost natural with minor imperfections
		\item 5 = Completely natural and human-like
	\end{itemize}
	
	\paragraph{2. Speaker Similarity (SMOS).}
	How closely does the synthesised voice resemble the reference speaker?
	Raters were asked to judge whether the quality, pitch, and vocal characteristics of the synthesised voice matched those of the reference recording.
	\begin{itemize}
		\item 1 = Completely different, like another person
		\item 2 = Somewhat different, low similarity
		\item 3 = Moderate similarity
		\item 4 = Very similar with minor differences
		\item 5 = Identical to the reference speaker
	\end{itemize}
	
\paragraph{3. Intelligibility (Intel.\ MOS).}
How well does the synthesised audio match the written text?
Raters were asked to note whether any words were missing, added, or mispronounced. Because Persian-specific phenomena such as Ezafe, unwritten short vowels, and context-dependent homograph pronunciation manifest as pronunciation errors to native listeners, these phenomena are also reflected in the intelligibility ratings. This scale used 0.5-point increments to permit finer-grained assessment of intelligibility:
\begin{itemize}
	\item 1 = Completely mismatched with the text
	\item 2 = Some correspondence but many errors
	\item 3 = Acceptable correspondence
	\item 4 = Good correspondence with minor errors
	\item 5 = Perfectly and exactly matched
\end{itemize}

\begin{figure}[h]
		\centering
		\includegraphics[width=0.80\columnwidth]{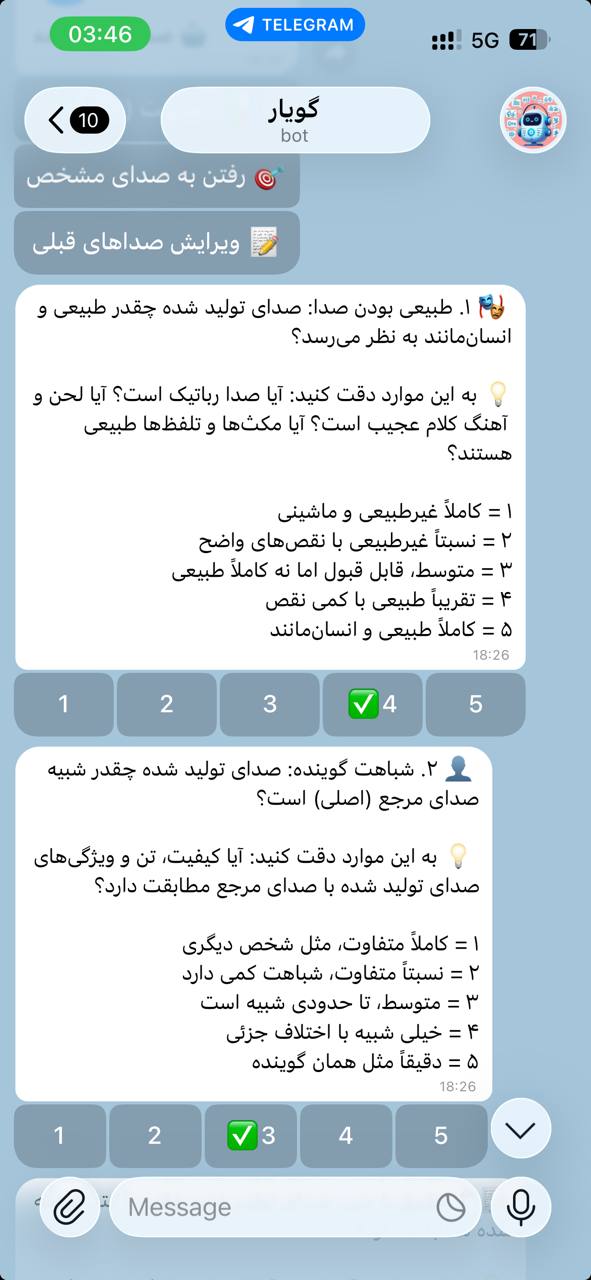}
		\caption{Screenshot of the Goyar Telegram-based annotation interface.
			The bot presents the reference text, delivers the reference and synthesised audio
			clips in sequence, and collects naturalness, speaker similarity, and intelligibility
			ratings via inline response buttons.}
		\label{fig:telegram_interface}
	\end{figure}

	Figure~\ref{fig:telegram_interface} shows a screenshot of the Goyar interface as displayed to raters.
	
	\subsection{Raters}
	\label{app:rater_details}
	Subjective ratings were collected via the Telegram-based interface described above. A total of 42 native Persian-speaking participants contributed ratings. For the subjective results reported in the paper, we retained the 22 raters who evaluated at least 10 items, yielding 1,812 retained ratings from 1,915 unique rating events. Table~\ref{tab:rater_demo} summarises rater demographics, and Table~\ref{tab:mos_by_rater_gender} reports mean scores grouped by rater gender.
	Participants were adult volunteers recruited through university research groups and personal contacts; participation was voluntary and uncompensated, and raters were told their ratings would be used for academic research on Persian speech synthesis. Only self-reported age and gender were collected. This minimal-risk perceptual evaluation was exempt from formal ethics-review-board approval under our institution's guidelines.
	
	\begin{table}[h]
		\centering
		\caption{Retained rater demographics (22 raters).}
		\label{tab:rater_demo}
		\small
		\setlength{\tabcolsep}{6pt}
		\begin{tabular}{lc}
			\toprule
			\textbf{Property} & \textbf{Value} \\
			\midrule
			Female / Male      & 11 / 11 \\
			Mean age (years)   & 28.6 \\
			Age range          & 19--60 \\
			\bottomrule
		\end{tabular}
	\end{table}
	
	\begin{table}[h]
		\centering
		\caption{Mean scores by rater gender.}
		\label{tab:mos_by_rater_gender}
		\small
		\setlength{\tabcolsep}{5pt}
		\begin{tabular}{lccc}
			\toprule
			\textbf{Rater gender} & \textbf{MOS} & \textbf{SMOS} & \textbf{Intel. MOS} \\
			\midrule
			Female & 3.46 & 3.82 & 3.84 \\
			Male   & 3.76 & 4.30 & 4.26 \\
			\bottomrule
		\end{tabular}
	\end{table}
	
	\paragraph{Inter-Rater Agreement.}
	We measured listener agreement using Krippendorff's $\alpha$, which supports different numbers of ratings per item. Each item received a median of 7 ratings. Ordinal $\alpha$ was 0.418 for MOS, 0.146 for SMOS, and 0.368 for intelligibility. Since MOS is reported as an average across listeners, we also estimated the reliability of the averaged scores using a crossed item and rater random-effects model. At the observed average of 7 listeners per item, reliability was 0.819 for MOS, 0.591 for SMOS, and 0.808 for intelligibility.
	
	\subsection{Reference Speakers}
	\label{app:ref_speakers_detail}
	The 90 synthesised samples correspond to 42 unique reference speakers stratified from Persian Common Voice~\cite{ardila2020common}, unseen during training. Speakers were stratified by gender and age group to ensure demographic coverage. Table~\ref{tab:spk_demo} summarises their demographics. Female reference speakers received marginally higher scores across all three dimensions (MOS 3.69 vs.\ 3.53, SMOS 4.13 vs.\ 3.99, intelligibility MOS 4.14 vs.\ 3.94).
	
	\begin{table}[h]
		\centering
		\caption{Reference speaker demographics (42 speakers).}
		\label{tab:spk_demo}
		\small
		\setlength{\tabcolsep}{6pt}
		\begin{tabular}{lcc}
			\toprule
			\textbf{Property} & \textbf{Count} & \textbf{\%} \\
			\midrule
			\multicolumn{3}{l}{\textit{Gender}} \\
			Female       & 17 & 40.5 \\
			Male         & 20 & 47.6 \\
			Not reported &  5 & 11.9 \\
			\midrule
			\multicolumn{3}{l}{\textit{Age group}} \\
			Teens        &  6 & 14.3 \\
			Twenties     & 10 & 23.8 \\
			Thirties     & 10 & 23.8 \\
			Forties      &  7 & 16.7 \\
			Fifties      &  1 &  2.4 \\
			Sixties      &  1 &  2.4 \\
			Not reported &  7 & 16.7 \\
			\bottomrule
		\end{tabular}
	\end{table}

\subsection{Human Verification of Transcripts}
\label{app:transcript_audit}

To evaluate ASR transcript accuracy, we randomly sampled 500 segments from the released TTS-ready subset, comprising 7,013 words. This evaluation therefore reflects the quality of retained segments and does not include rejected segments. Two native Persian speakers independently transcribed all 500 segments by listening. They agreed on 100\% of the words. Compared with these human reference transcriptions, the released ParsVoice transcripts achieved 4.90\% WER and 1.81\% CER, with 345 of 500 segments (69\%) transcribed perfectly. Before computing WER and CER, reference and hypothesis texts were normalised consistently by removing punctuation and diacritics and normalising ZWNJ/half-spaces, Arabic-to-Persian characters, digits, and whitespace.

The 90 synthesised evaluation samples comprise 942 words. The 4.78\% WER reported in Section~\ref{sec:tts_eval} corresponds to 45 word errors, with 52 of 90 samples (57.8\%) transcribed perfectly. The two transcribers differed on only 20 of the 942 words, corresponding to 97.8\% word-level agreement, indicating strong consistency between the independent human transcriptions.

\paragraph{Transcription protocol.}
Both transcribers worked independently and blind: they were shown neither the
ASR output nor, for the synthesised samples, the target text, and they did not
see each other's transcriptions. They were instructed to transcribe verbatim
exactly what they heard, including inserted, repeated, or hallucinated words,
rather than what they expected the sentence to be. Because Persian does not
write short vowels, transcribers were told to listen to the realised
pronunciation and to write the word actually produced; when a short vowel or an
Ezafe was pronounced differently from the intended word, the produced word was
transcribed as heard and therefore counted as a substitution.

% preamble: \usepackage{pgfplots}  \pgfplotsset{compat=1.18}  \usepgfplotslibrary{groupplots}

\definecolor{palBlue}{HTML}{2563EB}       % Full   (ParsVoice)
\definecolor{palAmber}{HTML}{D97706}      % 530 h  (FarsSpon)
\definecolor{palGreen}{HTML}{059669}      % 120 h  (Mana TTS)
\definecolor{palBlueDark}{HTML}{1E3A8A}
\definecolor{palAmberDark}{HTML}{92400E}
\definecolor{palGreenDark}{HTML}{064E3B}
\definecolor{palGrid}{HTML}{E2E8F0}
\definecolor{palAxis}{HTML}{CBD5E1}
\definecolor{palInk}{HTML}{0F172A}
\definecolor{palMuted}{HTML}{64748B}

\colorlet{lineBlue}{palBlue!55}           % raise/lower the number to taste
\colorlet{lineAmber}{palAmber!55}
\colorlet{lineGreen}{palGreen!55}

\begin{figure}[h]
	\centering
	\begin{tikzpicture}
		\begin{groupplot}[
			group style={group size=2 by 1, horizontal sep=1.5cm},
			width=0.5\linewidth, height=5cm,
			grid=major, grid style={palGrid, line width=0.5pt},
			axis line style={palAxis, line width=0.6pt},
			tick style={palAxis, line width=0.6pt},
			tick label style={font=\small, color=palMuted},
			label style={font=\normalsize, color=palInk},
			title style={font=\normalsize, color=palInk, yshift=-1mm},
			legend style={font=\small, draw=palAxis, fill=white,
				rounded corners=2pt, at={(0.97,0.97)}, anchor=north east},
			legend cell align=left,
			every axis plot/.append style={line width=1.4pt}]
			
			\nextgroupplot[xlabel={Epoch}, ylabel={Validation loss}, xmin=0, xmax=28,
			title={(a) equal epochs}]
			\addplot[lineGreen,mark=none] coordinates {(1,4.8635)(3,4.5358)(5,4.3325)(7,4.1944)(9,4.1017)(11,4.0140)(13,3.9570)(15,3.8939)(17,3.8460)(19,3.8079)(21,3.7773)(23,3.7525)(25,3.7260)(27,3.6982)};
			\addlegendentry{\textcolor{palGreenDark}{120\,h}}
			\addplot[lineAmber,mark=none] coordinates {(1,4.3894)(3,3.9434)(5,3.7673)(7,3.6630)(9,3.6081)(11,3.5618)(13,3.5297)(15,3.5061)(17,3.4903)(19,3.4702)(21,3.4592)(23,3.4471)(25,3.4354)(27,3.4229)};
			\addlegendentry{\textcolor{palAmberDark}{530\,h}}
			\addplot[lineBlue,mark=none]  coordinates {(1,3.8657)(3,3.5439)(5,3.4447)(7,3.3916)(9,3.3554)(11.5,3.3191)(13.5,3.2992)(15.5,3.2812)(17.5,3.2654)(19.5,3.2497)(21.5,3.2352)(23.5,3.2249)(25.5,3.2163)(26.5,3.2119)};
			\addlegendentry{\textcolor{palBlueDark}{Full}}
			
			\nextgroupplot[xlabel={Optimization steps (k)}, ymin=3.30, ymax=3.70,
			xmin=0, xmax=180, title={(b) equal steps}]
			\addplot[lineGreen,mark=none] coordinates {(35.4,3.6564)(46.8,3.5980)(58.2,3.5570)(69.7,3.5240)(81.0,3.4948)(92.5,3.4824)(103.9,3.4701)(115.3,3.4594)(126.8,3.4485)(138.2,3.4413)(149.6,3.4308)(157.6,3.4246)(162.2,3.4338)};
			\addplot[lineAmber,mark=none] coordinates {(35.4,3.6630)(45.5,3.6081)(55.6,3.5618)(65.8,3.5297)(75.9,3.5061)(86.0,3.4903)(96.2,3.4702)(106.3,3.4592)(116.4,3.4471)(126.5,3.4354)(136.7,3.4229)};
			\addplot[lineBlue,mark=none]  coordinates {(57,3.5439)(76,3.4850)(95,3.4447)(114.1,3.4124)(133.1,3.3916)(152.1,3.3680)(171.2,3.3554)};
		\end{groupplot}
	\end{tikzpicture}
	\caption{Validation loss.}
	\label{fig:loss}
\end{figure}

\section{Data-Scale Ablation}
\label{app:scale_ablation}

We fine-tuned XTTSv2 on nested 120\,h and 530\,h subsets of ParsVoice and on the full 2,200\,h training set using the same tokenizer, model configuration, and optimization settings. The subsets were constructed from a single random ordering of the full training corpus and truncated by cumulative duration, ensuring that each smaller subset is contained in the larger one: $120,\mathrm{h} \subset 530,\mathrm{h} \subset 2{,}200,\mathrm{h}$. In addition to WER and CER, we report NISQA-TTS~\cite{mittag20_interspeech}, a reference-free predictor of synthetic-speech naturalness.

\begin{table}[h]
	\centering
	\caption{Data-scale ablation after the same number of training epochs,
		scored with two independent ASR systems.}
	\label{tab:scale_ablation_epochs}
	\scriptsize
	\setlength{\tabcolsep}{4pt}
	\begin{tabular}{lccccc}
		\toprule
		& \multicolumn{2}{c}{\textbf{Google fa-IR}} &
		\multicolumn{2}{c}{\textbf{Whisper-fa}} & \\
		\cmidrule(lr){2-3}\cmidrule(lr){4-5}
		\textbf{Training data} & \textbf{WER (\%)} & \textbf{CER (\%)} &
		\textbf{WER (\%)} & \textbf{CER (\%)} & \textbf{NISQA} \\
		\midrule
		120\,h & 42.89 & 33.68 & 48.51 & 27.53 & 2.968 \\
		530\,h & 20.06 & 13.84 & \textbf{27.81} & \textbf{13.66} & 3.081 \\
		Full   & \textbf{17.83} & \textbf{12.14} & 30.36 & 14.58 & \textbf{3.165} \\
		\bottomrule
	\end{tabular}
\end{table}

Under equal epochs, performance generally improves as the training set grows, but this comparison also reflects optimization exposure because the same number of epochs corresponds to substantially different numbers of parameter updates across training-set sizes. Moreover, none of the three runs has clearly converged by 26 epochs, as validation loss is still decreasing for all conditions (Fig.~\ref{fig:loss}a).

We therefore also compare checkpoints at approximately matched optimization budgets. Under this setting, differences across data scales are smaller and no model is consistently best across all metrics. The runs also correspond to very different numbers of passes through their respective datasets: approximately 160k updates correspond to about 140 epochs for the 120\,h subset but only about 8.9 epochs for the full training set. Consistent with this difference, the 120\,h run begins to show signs of overfitting at high step counts, whereas the full-data run remains at an earlier stage of training and may benefit from additional optimization (Fig.~\ref{fig:loss}b).

\begin{table}[h]
	\centering
	\caption{Data-scale ablation at comparable optimization budgets,
		scored with two independent ASR systems.}
	\label{tab:scale_ablation_steps}
	\scriptsize
	\setlength{\tabcolsep}{4pt}
	\begin{tabular}{lccccc}
		\toprule
		& \multicolumn{2}{c}{\textbf{Google fa-IR}} &
		\multicolumn{2}{c}{\textbf{Whisper-fa}} & \\
		\cmidrule(lr){2-3}\cmidrule(lr){4-5}
		\textbf{Training data} & \textbf{WER (\%)} & \textbf{CER (\%)} &
		\textbf{WER (\%)} & \textbf{CER (\%)} & \textbf{NISQA} \\
		\midrule
		120\,h & \textbf{15.18} & \textbf{10.34} & 26.01 & 12.30 & 3.035 \\
		530\,h & 19.11 & 12.26 & 28.24 & 12.70 & \textbf{3.075} \\
		Full   & 17.62 & 11.96 & \textbf{25.58} & \textbf{10.90} & 3.015 \\
		\bottomrule
	\end{tabular}
\end{table}

\end{document}